\documentclass[11pt]{article}

\usepackage[latin1]{inputenc}

\usepackage{amstext,amsmath,amssymb,amsfonts,bbm,amsmath}
\usepackage{extarrows}
\usepackage{xcolor} 
\usepackage{hyperref}
\hypersetup{
    colorlinks=false,
    menubordercolor=red,
    linkbordercolor=blue
}
\usepackage{authblk}
\usepackage{caption} 
\usepackage{subcaption}
\usepackage{epsfig} 
\usepackage{color} 
\usepackage{graphicx} 
\usepackage{floatrow}
\usepackage{braket} 
\usepackage{slashed} 
\usepackage{dsfont} 
\usepackage{amsthm}  
\usepackage{multirow}

\usepackage{cite}

\allowdisplaybreaks[4]

\usepackage{psfrag}
\usepackage{tikz} 
\usetikzlibrary{arrows}
\usetikzlibrary{plotmarks}
\usetikzlibrary{external}
\usetikzlibrary{patterns}
\usepackage{pgfplots}
\pgfplotsset{compat=1.9}
\usetikzlibrary{patterns}
\usetikzlibrary{calc}
\usepackage[compat=1.1.0]{tikz-feynman}
\usetikzlibrary{automata,positioning}

\usepackage{float}  
\usepackage[lmargin=50pt,rmargin=60pt,tmargin=60pt,bmargin=65pt]{geometry}

\newcommand{\be}{\begin{equation}}
\newcommand{\ee}{\end{equation}} 
\newcommand{\nn}{\nonumber}

\renewcommand{\vec}[1]{\mathbf{#1}}

     \let\X=F

\newcommand{\CO}{\mathcal {O}}

\newcommand{\gt}{\tilde{g}}

\allowdisplaybreaks[4]

\numberwithin{equation}{section}

\theoremstyle{remark}

\begin{document}

\title{\bf Boundaries and Interfaces with Localized Cubic Interactions
in the $O(N)$ Model
}

\author[1]{Sabine Harribey}
\author[2,3]{Igor R. Klebanov}
\author[2,4]{Zimo Sun}

\affil[1]{\normalsize \it 
 NORDITA, Stockholm University and KTH Royal Institute of Technology,\authorcr Hannes Alfv\'{e}ns v\"{a}g 12, SE-106 91 Stockholm, Sweden
  \authorcr \hfill}

\affil[2]{\normalsize\it 
Department of Physics, Princeton University, Princeton, NJ 08544, USA
 \authorcr \hfill}

\affil[3]{\normalsize\it 
Princeton Center for Theoretical Science, Princeton, NJ 08544, USA
 \authorcr \hfill }
 
 \affil[4]{\normalsize\it 
Princeton Gravity Initiative, Princeton University, Princeton, NJ 08544, USA
 \authorcr \hfill }

\date{}

\maketitle

\hrule\bigskip

\begin{abstract}
We explore a new approach to boundaries and interfaces in the $O(N)$ model where we add certain localized cubic interactions. 
These operators are nearly marginal when the bulk dimension is $4-\epsilon$, and they explicitly break the $O(N)$ symmetry of the bulk theory down to $O(N-1)$.
We show that the one-loop beta functions of the cubic couplings are affected by the quartic bulk interactions. For the interfaces, we find real fixed points up to the critical value $N_{\rm crit}\approx 7$, while for $N> 4$ there are IR stable fixed points with purely imaginary values of the cubic couplings. For the boundaries, there are real fixed points for all $N$, but we don't find any purely imaginary fixed points. 
We also consider the theories of $M$ pairs of symplectic fermions and one real scalar, which have quartic $OSp(1|2M)$ invariant interactions in the bulk. We then add 
the $Sp(2M)$ invariant localized cubic interactions.
The beta functions for these theories are related to those in the $O(N)$ model via the replacement of $N$ by $1- 2M$. In the special case $M=1$, there are boundary or interface fixed points that preserve the 
$OSp(1|2)$ symmetry, as well as other fixed points that break it.

\end{abstract}

\hrule\bigskip

\tableofcontents

\section{Introduction and summary}
\label{sec:introduction}
The critical $O(N)$ vector models are arguably the most thoroughly studied class of 3D Conformal Field Theories (CFTs). They can be described by Euclidean field theory of $N$ scalar fields $\phi_I$ with the quartic $O(N)$ invariant interactions. While these CFTs do not appear to be exactly solvable, there is a variety of approximation methods available for them, including the $4-\epsilon$ expansion \cite{Wilson:1971dc} and the Conformal Bootstrap \cite{Polyakov:1974gs,Kos:2013tga} (for excellent reviews, old and new, see \cite{Wilson:1973jj,Poland:2018epd,Henriksson:2022rnm}). Another useful tool is the $1/N$ expansion, which can be carried out in continuous dimension $D$ \cite{Vasiliev:1981dg,Moshe:2003xn}.
This expansion is related \cite{Klebanov:2002ja} via the AdS/CFT correspondence to the higher-spin quantum gravity in the $D+1$ dimensional Anti-de Sitter space 
\cite{Vasiliev:1990en,Vasiliev:2003ev}.

It is of obvious interest to study the $O(N)$ models on spaces with boundaries, as well as the closely related problem of introducing interfaces, i.e.  
codimension-one defects. Research on the critical behavior in such systems dates back many years \cite{AJBray_1977,Ohno:1983lma,PhysRevB.30.300,gompper1985conformal,McAvity:1995zd} and is reviewed in \cite{Diehl:1996kd}. 
More recent results on the various boundary universality classes (called ``special", ``ordinary", and ``extraordinary") were obtained in \cite{Liendo:2012hy,Giombi:2020rmc}.
In particular, the ``extraordinary" critical interfaces and boundaries break the $O(N)$ symmetry to $O(N-1)$. For $N>1$ and $D>3$, 
this universality class is believed to be equivalent to the ``normal" universality class obtained via explicit symmetry breaking on the boundary \cite{Diehl:1996kd}.
While the extraordinary universality class for boundaries or interfaces was known to exist in bulk dimension greater than $3$, it was not completely clear what happens to it for $D=3$. During the past three years, this problem was revisited in papers \cite{Metlitski:2020cqy,Padayasi:2021sik,Toldin:2021kun,Krishnan:2023cff}. Using a combination of Renormalization Group (RG) analysis and $1/N$ expansion, they demonstrated the existence of the 3D ``extraordinary log" universality class. For boundaries, this class exists only for $N$ smaller than a critical value which is above $3$
\cite{Metlitski:2020cqy,Padayasi:2021sik,Toldin:2021kun}, but for the interfaces it appears to exist for all $N$ \cite{Krishnan:2023cff}.

In recent papers \cite{Trepanier:2023tvb,Giombi:2023dqs,Raviv-Moshe:2023yvq,Cuomo:2023qvp}, another approach to the surface defect was used
where the bulk was taken to have continuous dimension $D$. 
In $D=4-\epsilon$, the quadratic operators on the defect are nearly marginal, so that the defect beta function can be calculated perturbatively. 
The papers \cite{Trepanier:2023tvb,Giombi:2023dqs} also contain perturbative analyses in $D=6-\epsilon$ where the bulk $O(N)$ model is defined by the cubic action \cite{Fei:2014yja}. Here
the surface defects, either ordinary or extraordinary, correspond to turning on the nearly marginal operators that are linear in the fields. 
The papers \cite{Trepanier:2023tvb,Giombi:2023dqs,Raviv-Moshe:2023yvq} also studied the $1/N$ expansion in presence of a surface defect and found that it becomes singular as $D$ approaches $3$ from above.

In this paper, we will use a different way to formulate a symmetry breaking defect via the $4-\epsilon$ expansion: instead of $2$, we will take the defect dimension to be 
$d=D-1=3-\epsilon$, so that it is always of codimension $1$.
We will turn on {\it cubic} interactions localized on the interface or boundary, which are nearly marginal for $D=4-\epsilon$. It is not hard to see that they can preserve at most the $O(N-1)$ symmetry, in which case they assume the form \cite{Fei:2014yja}
\begin{equation}
\frac{1}{2} \lambda_1 \phi_N \sum_{a=1}^{N-1} \phi_a \phi_a+ \frac{1}{6} \lambda_2 \phi_N^3 \, .
\end{equation} 
The explicit breaking of $O(N)$ to $O(N-1)$ by the cubic terms is analogous to the explicit breaking by the linear term on the extraordinary surface defect in $D=6-\epsilon$
 \cite{Trepanier:2023tvb,Giombi:2023dqs,Giombi:2020rmc}. The standard formulation of the normal transition in $D=4-\epsilon$ includes such a linear term, but we have fine-tuned it, as well as some other relevant operators, to zero. Thus, our approach appears to describe a multicritical version of the normal (extraordinary) transition.

We will derive the beta functions for the interface coupling constants $\lambda_1$ and $\lambda_2$ that include the effects of the quartic bulk interactions.\footnote{A codimension-one defect or boundary with the nearly marginal quartic interactions can be introduced into the $3-\epsilon$ dimensional $O(N)$ model with sextic interactions \cite{Herzog:2020lel,SoderbergRousu:2023vpe}.} 
Our analysis of interface fixed points in $D=4-\epsilon$ reveals the existence of real fixed points only up to a critical value of $N$. We find $N_{\rm crit}\approx 7.1274$,
which is similar to the upper critical value $N_{\rm crit}=10$ found in \cite{Trepanier:2023tvb,Giombi:2023dqs,Raviv-Moshe:2023yvq} for the extraordinary surface defect.
For $N> 4$ there are IR stable fixed points with purely imaginary values of the cubic couplings $\lambda_1, \lambda_2$.
In this case, the path integral appears to be well defined, but the theories are not expected to be unitary.
Interestingly, for the boundary case, where the beta functions are similar to the interface ones but have different coefficients, we find that there are real fixed points for all $N$.
In both the interface and the boundary cases, we find that the leading contribution to the VEV of the bulk field $\phi_N$ arises at the two-loop level and is of order $\epsilon^{3/2}$.

We also consider the theories of $M$ pairs of symplectic fermions and one real scalar with quartic $OSp(1|2M)$ invariant interactions in the bulk. They may be regarded as continuations of the $O(N)$ models to odd negative values of $N$, i.e. $N=1-2M$. Here we also find real fixed points for the cubic interactions localized on the interfaces or boundaries. In the special case $M=1$, we find that some fixed points preserve the $OSp(1|2)$ symmetry while others violate it. 

This paper is organized as follows. In section \ref{sec:vector model}, we define the model with cubic interactions on an interface and compute the corresponding one-loop beta functions. In section \ref{sec:epsilonexp}, we then study the fixed points of both free and interacting bulk theories in a $D=4-\epsilon$ expansion before computing in section \ref{quadO} the dimensions of quadratic operators at the fixed points. In section \ref{sec:symp}, we study the case of an interface in a model with symplectic fermions. Then, in section \ref{sec:boundary}, we compare the previous results with the case of cubic interactions on a boundary. Finally, some exotic large $N$ limits are discussed in appendix \ref{Exotic}.

\section{Cubic interactions localized at the interface}
\label{sec:vector model}

We consider the $O(N)$ vector model in $(d+1)$-dimensional Euclidean space. The coordinates of $\mathbb R^{d+1}$ are labeled by $x^\mu=(\vec x, y)$, where $\mu=1,2,\cdots, d+1$, and $\vec x$ is a vector in $\mathbb R^d$. Inserting an interface with cubic interactions at $y=0$, we obtain the action:
\begin{align} \label{eq:action_vectpr}
		S[\phi]  \, &= \,  \int d^{d+1}x \, \bigg[ \frac{1}{2} \partial_{\mu}\phi_I\partial^{\mu}\phi_I + \frac{\lambda_4}{4!}\left(\phi_I \phi_I \right)^2 \bigg]   +  \int d^{d}\vec x \bigg[
		\frac{\lambda_1}{2} \, 
		\phi_N \phi_a \phi_a +\frac{\lambda_2}{3!}\phi_N^3 \bigg] \, ,
\end{align}
where the index $I$ is summed from $1$ to $N$ while the index $a$ is summed from $1$ to $N-1$. The $O(N-1)$ invariant cubic interactions on the interface have the form introduced in
 \cite{Fei:2014yja}.
A reason to include the cubic interactions on the interface is that they are marginal for $d=3$, just like the quartic bulk interactions. 
Therefore, in $d=3-\epsilon$ the coupled bulk-interface system may be studied perturbatively.\footnote{Perhaps we can also view the cubic terms on the interface as resulting from giving an expectation value for $\phi_N$ on the interface. Then, after expanding around the vacuum where $\phi_N \sim \delta(y)$, we find the $O(N-1)$ invariant localized cubic terms like those in \eqref{eq:action_vectpr}.}

The bulk propagator of the free theory is given by 
\begin{align}\label{interprop}
\langle\phi_I(x_1)\phi_J(x_2) \rangle=\delta_{IJ}\int\,\frac{d^{d+1} p}{(2\pi)^{d+1}}\frac{e^{i p\cdot x_{12}}}{p^2} =\delta_{IJ}\frac{C_\phi}{|x_{12}|^{d-1}},\,\,\,\,\,  C_\phi=\frac{\Gamma\left(\frac{d-1}{2}\right)}{4\pi^{\frac{d+1}{2}}}~,
\end{align}
where $x_{12}^\mu\equiv x_1^\mu- x_2^\mu$. Performing Fourier transformation of the free propagator along the interface directions yields 
\begin{align}\label{intermomentum}
\langle\phi_I(\vec p_1, y_1)\phi_J(\vec p_2, y_2) \rangle &=\delta_{IJ}C_\phi \int d^d \vec x_1\,d^d\vec x_2 \frac{e^{i \vec p_1\cdot \vec x_1+i \vec p_2\cdot \vec x_2}}
{\left ( \vec x_{12}^2+y_{12}^2 \right )^{\frac{d-1}{2}}}\nn\\
&=
\frac{\delta_{IJ}}{4\pi^{\frac{d+1}{2}}}\int_0^\infty\frac{ds}{s}s^{\frac{d-1}{2}} e^{-s y_{12}^2}\, \int\, d^d \vec x_1\, d^d\vec x_2 \, e^{-s \vec x_{12}^2+i \vec p_1\cdot \vec x_1+i \vec p_2\cdot \vec x_2}\nonumber\\
&=\frac{(2\pi)^d\delta^d(\vec p_1\!+\!\vec p_2)\delta_{IJ}}{4\,\sqrt{\pi}}\int_0^\infty\frac{ds}{s^{\frac{3}{2}}}\,e^{-s y_{12}^2-\frac{\vec p_1^2}{4s}}=(2\pi)^d\delta^d(\vec p_1\!+\!\vec p_2)\,\delta_{IJ} \frac{e^{-|\vec p_1||y_{12}|}}{2|\vec p_1|}~.
\end{align}
By choosing $y_1=0$ or $y_1=y_2=0$, we obtain the free interface-to-bulk propagator in the mixed space or free interface propagator in momentum space
\begin{align}
K_{IJ}(\vec p, y) = \frac{ e^{-|\vec p||y|}}{2|\vec p|} \delta_{IJ}, \,\,\,\,\,\,\, G_{IJ} (\vec p) = \frac{\delta_{IJ}}{2|\vec p|}~.
\end{align}

\subsection{One-loop renormalization}
\begin{figure}[h]
\centering
\begin{subfigure}[b]{0.5\textwidth}
\centering
\includegraphics[scale=2]{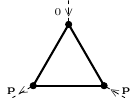}
\caption{$T$}
\label{TT}
\end{subfigure}
\hfill
\begin{subfigure}[b]{0.45\textwidth}
\centering
\includegraphics[scale=2]{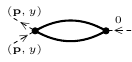}
\caption{$B$}
\label{BB}
\end{subfigure}
\caption{One-loop corrections to the cubic couplings. }
\label{fig:1_2_loops}
\end{figure}

To look for fixed points, we compute $\beta$ functions for all the couplings up to one loop. For the bulk quartic coupling $\lambda_4$, its $\beta$ function 
is not affected by the interface \cite{ZinnJustin:2002ru}:
\begin{align}\label{betaquartic}
\beta_{\lambda_4} = -\epsilon \lambda_4+\frac{N+8}{3}\frac{\lambda_4^2}{(4\pi)^2}~.
\end{align}
For the cubic couplings on the interface, there are two types of diagrams contributing to their one-loop renormalization, as shown in Figure~\ref{fig:1_2_loops}. The diagram $T$ only involves cubic couplings while the diagram $B$ involves both quartic and cubic couplings.

The one-loop corrections to $\lambda_1$ and $\lambda_2$ vertices are
\begin{align}
&\Gamma_1^{(1)} = -\lambda_1^2(\lambda_1+\lambda_2)I_T+\frac{\lambda_4}{6}\left((N+5)\lambda_1+\lambda_2\right)I_B-\delta\lambda_1\, , \nn\\
&\Gamma_2^{(1)} = -\left((N-1)\lambda_1^3+\lambda_2^3\right)I_T+\frac{\lambda_4}{2}\left((N-1)\lambda_1+3\lambda_2\right)I_B-\delta\lambda_2 \, ,
\end{align}
where 
\begin{align}\label{IT}
&I_T =\frac{\mu^\epsilon}{8} \int \frac{d^d \vec k}{(2\pi)^d} \frac{1}{|\vec k|^2 |\vec k+\vec p|} = \frac{1}{(4\pi)^2\,\epsilon}+\mathcal{O}(1) 
\end{align}
corresponds to the diagram $T$, and 
\begin{align}\label{IB}
&I_B =\mu^\epsilon\int\frac{d^d \vec k}{(2\pi)^d}\, \int_{\mathbb R} d y\, \left(\frac{1}{2|\vec k|}\right)^2 e^{-2 (|\vec p|+|\vec k|)|y|} = \frac{\mu^\epsilon}{4}\int\frac{d^d \vec k}{(2\pi)^d}\frac{1}{|\vec k|^2 (|\vec p|+|\vec k|)} =  \frac{2}{(4\pi)^2\,\epsilon}+\mathcal{O}(1) 
\end{align}
corresponds to the diagram $B$. The evaluation of $I_T$ is based on the formula 
\begin{equation}
\label{eq:bubble}
\int \frac{d^d \vec k}{(2\pi)^d}\frac{1}{|\vec k|^{2\alpha} |\vec k+\vec p|^{2\beta}}=\frac{1}{(4\pi)^{\frac{d}{2}}|\vec p|^{2\alpha+2\beta-d}}\frac{\Gamma(\frac{d}{2}-\alpha)\Gamma(\frac{d}{2}-\beta)\Gamma(\alpha+\beta-\frac{d}{2})}{\Gamma(\alpha)\Gamma(\beta)\Gamma(d-\alpha-\beta)} \,
\end{equation}
and the evaluation of $I_B$ can be done easily after writing it as a one-dimensional integral over $|\vec k|$. In the minimal subtraction scheme, we choose the counterterms to be 
\begin{align}
&\delta\lambda_1 = \frac{\lambda_4\left((N+5)\lambda_1+\lambda_2\right)}{3 (4\pi)^2\,\epsilon}-\frac{\lambda_1^2(\lambda_1+\lambda_2)}{(4\pi)^2\,\epsilon}~,\nn\\
&\delta\lambda_2 =\frac{\lambda_4\left((N-1)\lambda_1+3\lambda_2\right)}{(4\pi)^2\,\epsilon}-\frac{\left((N-1)\lambda_1^3+\lambda_2^3\right)}{(4\pi)^2\,\epsilon}~.
\end{align}
Requiring $\mu^{\frac{\epsilon}{2}} (\lambda_1+\delta\lambda_1)$ and $\mu^{\frac{\epsilon}{2}} (\lambda_2+\delta\lambda_2)$ to be $\mu$ independent\footnote{Here we use the fact that the wavefunction renormalization of $\phi_I$ starts with $\lambda_4^2$, and hence does not contribute at the one-loop order.}, we  obtain the $\beta$ functions of $\lambda_1$ and $\lambda_2$
\begin{align}\label{betacubic}
&\beta_{\lambda_1}=-\frac{\epsilon}{2}\lambda_1-\frac{\lambda_1^2(\lambda_1+\lambda_2)}{(4\pi)^2}+\frac{\lambda_4\left((N+5)\lambda_1+\lambda_2\right)}{3 (4\pi)^2}~,\nn\\
&\beta_{\lambda_2}=-\frac{\epsilon}{2}\lambda_2-\frac{\left((N-1)\lambda_1^3+\lambda_2^3\right)}{(4\pi)^2}+\frac{\lambda_4\left((N-1)\lambda_1+3\lambda_2\right)}{(4\pi)^2}~.
\end{align}
For the simplicity of notation, we make the following rescaling 
\begin{align}\label{gdef}
g_4 = \frac{\lambda_4}{(4\pi)^2} \, , \,\,\,\,\,\,  g_{1, 2} = \frac{\lambda_{1,2}}{2\pi}~,
\end{align}
and then the $\beta$ functions of these rescaled couplings can be summarized as 
\begin{align}\label{betag}
\beta_{g_4}&= -\epsilon g_4 +\frac{N+8}{3}g_4^2  \, , \crcr
\beta_{g_1}&=-\frac{\epsilon}{2}g_1 -\frac{1}{4}\,g_1^2\left(g_1+g_2\right) +\frac{1}{3}g_4\big ((N+5)g_1+g_2\big ) \, ,  \crcr
\beta_{g_2}&= -\frac{\epsilon}{2}g_2 -\frac{1}{4}\left((N-1)g_1^3+g_2^3\right) +g_4 \big ( (N-1)g_1+3g_2 \big )   \, .
\end{align}

\section{Fixed points in $D=4-\epsilon$}
\label{sec:epsilonexp}

\subsection{Free bulk}\label{freebulk}

We first consider the fixed points with no interaction in the bulk, $\lambda_4=0$ (this is a warm-up to the more interesting case where we have the interacting $O(N)$ model in the bulk, which will be discussed in the next section).
A similar model with no bulk interactions was considered in Appendix A.2 of \cite{Giombi:2019enr}, but with a boundary instead of an interface. The one-loop beta functions are:
\begin{align}
\label{freebulk}
\beta_{g_1}&=-\frac{\epsilon}{2}g_1 -\frac{1}{4}\,g_1^2\left(g_1+g_2\right) \, ,  \nn\\
\beta_{g_2}&= -\frac{\epsilon}{2}g_2 -\frac{1}{4}\left((N-1)g_1^3+g_2^3\right)   \, .
\end{align}
As usual for theories with cubic couplings, the solutions of the beta function equations come in pairs: $(g^\star_1,g^\star_2), (-g^\star_1,-g^\star_2)$. These fixed points are mapped into each other by $\phi_I\rightarrow - \phi_I$, hence they are equivalent.

The  $N=1$ theory is equivalent to setting $g_1=0$. Then $\beta_{g_2}$ reduces to $-\frac{\epsilon g_2}{2}  -\frac{g_2^3}{4}$, which has two non-trivial fixed points
\begin{equation}\label{N=1fix}
g_2^{\star}=\pm i\sqrt{2\,\epsilon}+\mathcal{O}(\epsilon^{3/2}) \, .
\end{equation}
They are purely imaginary, and have the critical exponent $\omega =\left.\frac{\partial\beta_{g_2}}{\partial g_2}\right|_{g_2^\star} =\epsilon$.
This result is very similar to the usual bulk Yang-Lee model where a stable purely imaginary fixed point is found in $D=6-\epsilon$ \cite{Fisher:1978pf}.

The $N=2$ case has more intricate structures. There are two pairs of non-trivial fixed points: 
\begin{align}
(g^\star_1,g^\star_2)=(0, \pm i\sqrt{2\epsilon}), \,\,\,\,\,\,\ (g_1^\star, g_2^\star) = \pm  (i \sqrt{\epsilon},i \sqrt{\epsilon})~.
\end{align}
The critical exponents\footnote{The critical exponents are defined as the eigenvalues of the $2\times 2$ matrix $\frac{\partial \beta_{g_j}}{\partial g_i}$ evaluated at fixed points.} for these two pairs of fixed points are respectively $(-\frac{\epsilon}{2},\epsilon)$ and $(0,\epsilon)$. In the first scenario, the $N=2$ theory contains two decoupled copies of the $N=1$ theory, at the trivial and non-trivial fixed points respectively.
In the second scenario,  the $N=2$ theory becomes the sum of two decoupled copies of the $N=1$ theory both at the non-trivial fixed point, because the cubic interaction is proportional to $\frac{\sqrt{2}\, g_2^\star}{6}\left(\phi_+^3+\phi_-^3\right)$ with $\phi_\pm = \frac{\phi_2\pm \phi_1}{\sqrt{2}}$. The critical exponent $\omega_+ = \epsilon$ corresponds to the slightly irrelevant cubic operator $\CO_+ = \phi_+^3+\phi_-^3$, which has scaling dimension $\Delta_+ = d+\omega_+ =3$. The other critical component $\omega_-= 0$ corresponds to the cubic  operator $\CO_- = \phi_+ \phi_-^2+\phi_-\phi_+^2$. It is marginal on the interface at the one-loop order. Its scaling dimension can be alternatively computed by noticing that 
\begin{align}\label{21rela}
\Delta_{\CO_-}^{N=2} = \Delta_{\phi}^{N=1} +\Delta_{\phi^2}^{N=1} 
\end{align}
where $\Delta_{\phi}^{N=1}$ and $\Delta_{\phi^2}^{N=1} $ denote the scaling dimensions of $\phi$ and $\phi^2$ of the $N=1$ theory at the non-trivial fixed points specified by \eqref{N=1fix}. At the one-loop  order $\phi$ is not renormalized, and hence $\Delta_{\phi}^{N=1} = \frac{d-1}{2}$. For $\phi^2$, its anomalous dimension is computed in section \ref{quadO}. More explicitly, $\gamma^{N=1}_{\phi^2} = -\frac{1}{4}\left(g_2^\star\right)^2 = \frac{\epsilon}{2}$. Substituting this into \eqref{21rela}, we obtain the scaling dimension of $\CO_-$, which indeed agrees with $d+\omega_- = 3-\epsilon$.

For $N>2$, the beta equations have nine solutions. One is the trivial fixed point which is unstable. Two of them are purely imaginary with $g_1=0$ and 
$g_2=\pm i\sqrt{2\epsilon}$,
and critical exponents $(-\frac{\epsilon}{2}, \epsilon)$. In this case the $N-1$ scalars $\phi_a$ are completely decoupled and free, and the scalar field $\phi_N$ has the form of the $N=1$ theory. 
The remaining six solutions have both couplings non-zero. We have one pair of real fixed points and two pairs of complex conjugate fixed points. So $N_{crit}=2$ is the lower bound for $N$ that admits non-trivial real fixed points. These non-trivial  real fixed points have one irrelevant and one relevant directions. The complex fixed points have one irrelevant direction and one complex critical exponent with positive real part.

We can analyze more precisely the solutions for both finite $N$  and large $N$ following the method of \cite{Fei:2014yja}.
We denote $g_1=\sqrt{8\,\epsilon}\,x$, $g_2=\sqrt{8\,\epsilon}\,y$ and we want to solve for both $x$ and $y$ non-zero. After some manipulation the vanishing of beta functions reduces to:
\begin{align}
& 4x (x+y)+1=0 \crcr
& -(N-1)x^3+x^2y+xy^2-y^3=0 \, .
\end{align}
The change of variable $y = z x$ effectively decouples the two equations
\begin{align}\label{xzeq}
& 4 x^2 (z+1)= -1 \crcr
&z^3-z^2-z +N-1=0~.
\end{align}
The discriminant of the cubic equation is $\Delta=-(N-2)(27N-22)$ 
which is zero when $N=2$ and strictly negative when $N>2$. For $N>2$, the cubic equation thus has one real root and two complex conjugate roots. To determine if the  real root  leads  to real or purely imaginary fixed points, we rewrite the cubic equation of $z$ as $(z+1)(z-1)^2 = 2-N$, which immediately implies that the real solution  satisfies $z+1<0$ when $N>2$. Therefore, $x$ and $y$ are also real at this point. In addition, the relations \eqref{xzeq} allow us to express $\frac{\partial \beta_{g_j}}{\partial g_i}$ only in terms of $z$ at fixed points
\begin{align}
\frac{\partial \beta_{g_j}}{\partial g_i} =\frac{\epsilon}{2(z+1)} \left(
\begin{array}{cc}
 z+2 & 1 \\
 3 z(z+1-z^2) & 3 z^2-(z+1) \\
\end{array}
\right)~,
\end{align}
which leads to the following critical exponents at the one-loop level
\begin{align}\label{omegapmfree}
\omega_+ = \epsilon, \,\,\,\,\, \omega_- =  \frac{(z-1) (3 z+1)}{2 (z+1)}\epsilon~.
\end{align}
The first critical exponent implies that there is always a slightly irrelevant operator $\CO_+$ of scaling dimension $\Delta_+=3$ for any $N$ at all fixed points. The second critical exponent is negative at the real fixed points since $z<-1$, corresponding to a relevant operator, and complex at the complex fixed points, satisfying $\text{Re} (\omega_-)>0$.

\subsection{Interacting bulk}\label{Intint}

For the interacting theory in the bulk, we tune the quartic coupling to the usual Wilson-Fisher fixed point in $4-\epsilon$ dimension:
\begin{equation}
g_4^{\star}=\frac{3\epsilon}{N+8}+\mathcal{O}(\epsilon^2) \, .
\label{eq:quarticWF}
\end{equation}
Substituting this value into the other one-loop beta functions, we find that they become
\begin{align}
\beta_{g_1}&=\frac{\epsilon (N+2)}{2 (N+8)}g_1 +\frac{\epsilon}{N+8} g_2  -\frac{1}{4}\,g_1^2\left(g_1+g_2\right) \, ,  \nn\\
\beta_{g_2}&= \frac{\epsilon (10- N)}{2 (N+8)}g_2 +\frac{3\epsilon (N-1)}{N+8} g_1 -\frac{1}{4}\big ((N-1)g_1^3+g_2^3 \big)   \, .
\end{align}
This is different from the beta functions \eqref{freebulk} for the case of the free bulk, and therefore the structure of fixed points is quite different.

When $N=1$, this leads to the following beta function for $g_2$
\begin{align}
N=1: \,\,\,\beta_{g_2}= \frac{\epsilon}{2} g_2- \frac{1}{4}g_2^3~.
\end{align}
In contrast to the free case, $\beta_{g_2}$ now has a pair of {\it real} non-trivial fixed points for the cubic coupling $g_2$:
\begin{equation}\label{g2int}
g_2^{\star}=\pm \sqrt{2\,\epsilon}+\mathcal{O}(\epsilon^{3/2}) \, .
\end{equation}
Thus, unlike in the free bulk case, the theory at the fixed point looks unitary.
They have the critical exponent $\omega_2=-\epsilon$, which means that the cubic operator $\phi^3$ is slightly relevant with scaling dimension $\Delta_{\phi^3} = 3-2\epsilon$ at these fixed points.

To solve for the fixed points for $N\ge 2$, we make the same substitution as before, i.e. $g_1=\sqrt{8\,\epsilon} \, x, g_2=\sqrt{8\,\epsilon}\, y$, which yields
\begin{align}\label{xyint}
&N x+2(x+ y)=4 ( N+8) x^2(x+ y), \nn\\
&6 (N-1) x-(N-10) y =4 (N+8)\left((N-1)x^3+y^3\right)~.
\end{align}
From \eqref{xyint}, we obtain a quartic equation of $z = \frac{y}{x}$ 
\begin{align}\label{z4}
P_{N}(z)\equiv 2 z^4+(N+2) z^3+(N-10) z^2-3 (N+2) z+(N-1)(N-4)=0~,
\end{align}
and $x$ is related to $z$ by 
\begin{align}\label{ztox2}
x^2 = \frac{1}{4(N+8)} \left(\frac{N}{1+z}+2\right)~.
\end{align}

\begin{figure}[t]
\centering
\includegraphics[width=8cm]{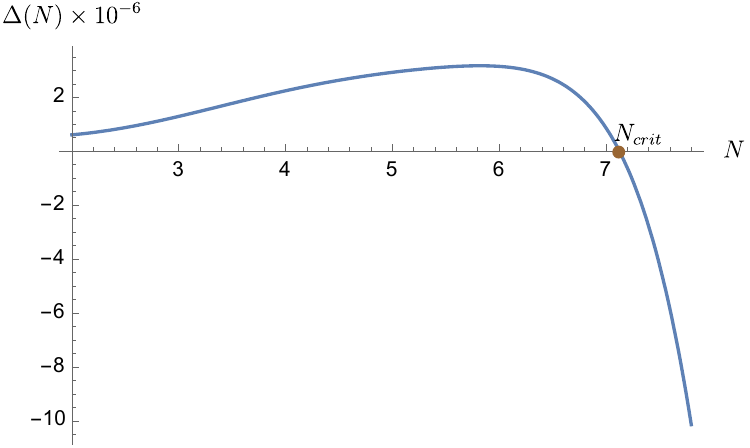}
\caption{The discriminant (rescaled by $10^{-6}$) of the quartic equation \eqref{z4}. It vanishes at the critical value $N_{\rm crit}\approx 7.1274 $.}
    \label{fig:disc}
\end{figure}

We plot the discriminant $\Delta(N)$ of the quartic polynomial $P_N(z)$ in Figure~\ref{fig:disc}. $\Delta(N)$ is positive for $2\le N\le 7$, meaning that \eqref{z4} has four different real roots in this region. It is negative for $N\ge 8$ and hence \eqref{z4} has two real roots and two complex roots that are conjugate of each other. Increasing $N$ from 7 to 8, two out of the four real roots of $P_N(z)$ collide as $N$ approaches the critical value $N_{\rm crit}\approx 7.1274$ where the discriminant vanishes, and subsequently go off to the complex $z$ plane. 

In general, each complex $z$ gives rise to  a pair of complex fixed points, and each real $z$ gives rise to either a pair of real fixed points or a pair of purely imaginary fixed points, depending on the sign of $\frac{N}{1+z}+2$, c.f. \eqref{ztox2}. There exists another critical value $N'_{crit}$ of $N$, such that $\frac{N}{1+z}+2<0$ when $N>N'_{crit}$ and $\frac{N}{1+z}+2>0$ when $N<N'_{crit}$, at some real root $z$ of $P_N(z)=0$. The critical value $N'_{crit}$ is fixed by requiring 
\begin{align}
P_{N'_{crit}} \left(-\frac{N_{crit}'+2}{2}\right)=\frac{1}{4}N'_{crit}(N'_{crit}-4)(N'_{crit}+8)=0 \, .
\end{align}
The relevant solution for us is $N'_{crit}=4$. When $N$ approaches 4 from below, a pair of real fixed points collide  and annihilate each other. After $N$ crosses 4, the two real fixed points reappear with purely imaginary $x$ and $y$.
In total,  for a generic positive integer $N$ the beta functions have nine fixed points including the trivial one, except two special cases, i.e. $N=4$ and $N=2$. The former, as just discussed, corresponds to a pair of real fixed points merging into the trivial one.  In the latter case, \eqref{z4} has the solution $z=-1$, or equivalently $x=-y$. Plugging it into the first line of \eqref{xyint} then yields the trivial fixed point $x=y=0$. Altogether, when $N=2$ or $4$, there are only seven fixed points in total. 

\begin{figure}[h]
     \centering
     \begin{subfigure}[b]{0.3\textwidth}
         \centering
         \includegraphics[width=\textwidth]{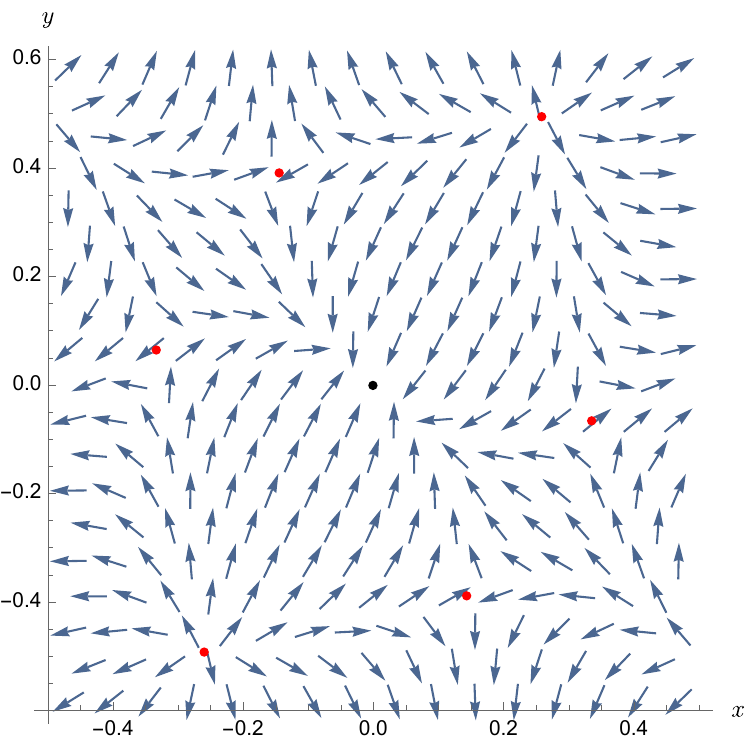}
         \caption{$N=2$}
         \label{fig:N=2}
     \end{subfigure}
     \hfill
     \begin{subfigure}[b]{0.3\textwidth}
         \centering
         \includegraphics[width=\textwidth]{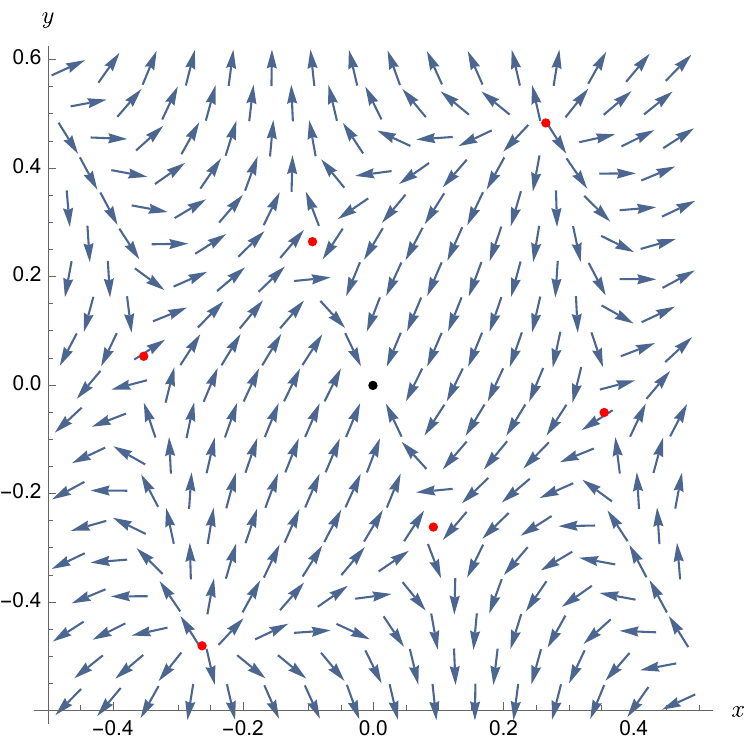}
         \caption{$N=3$}
         \label{fig:N=3}
     \end{subfigure}
     \hfill
     \begin{subfigure}[b]{0.3\textwidth}
         \centering
         \includegraphics[width=\textwidth]{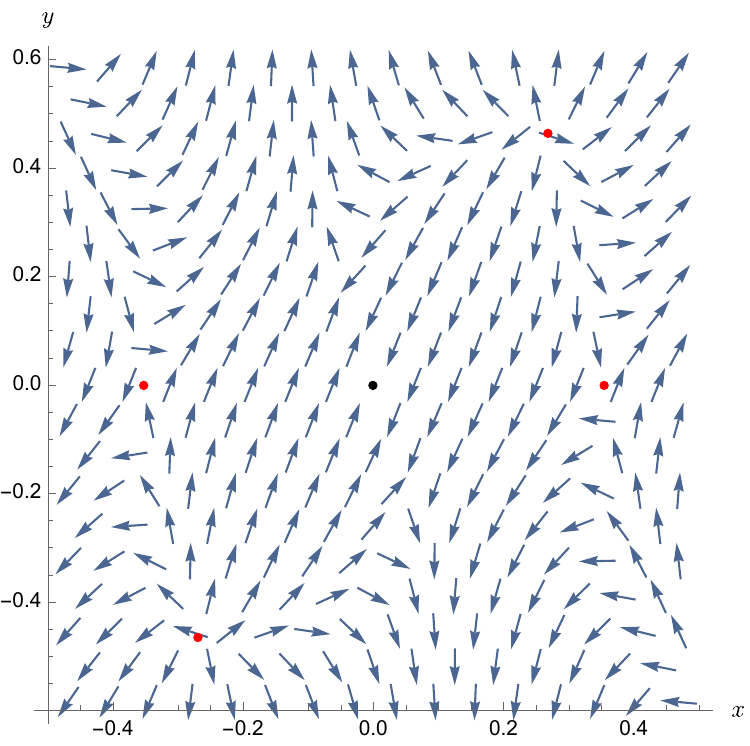}
         \caption{$N=4$}
         \label{fig:N=4}
     \end{subfigure}
          \begin{subfigure}[b]{0.3\textwidth}
         \centering
         \includegraphics[width=\textwidth]{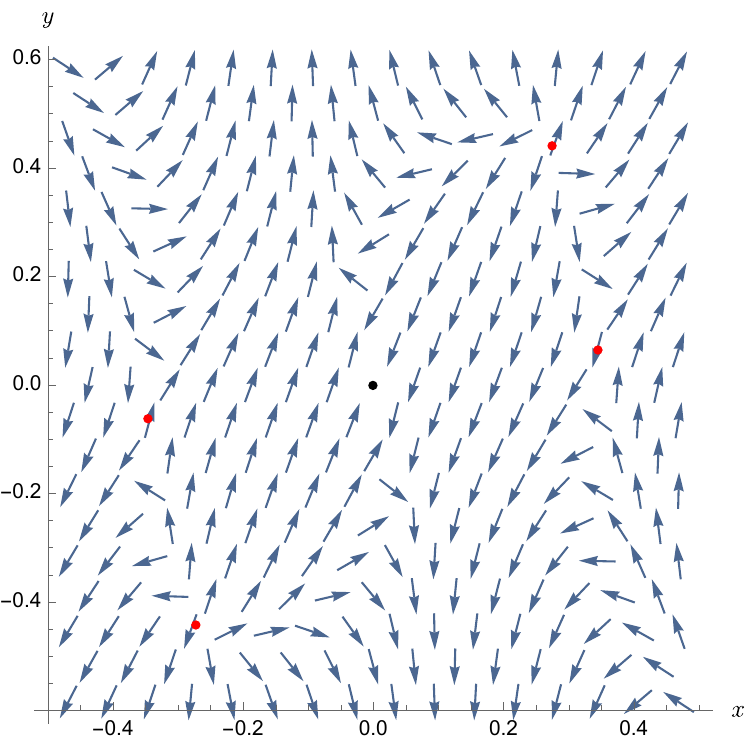}
         \caption{$N=5$}
         \label{fig:N=5}
     \end{subfigure}
     \hfill
     \begin{subfigure}[b]{0.3\textwidth}
         \centering
         \includegraphics[width=\textwidth]{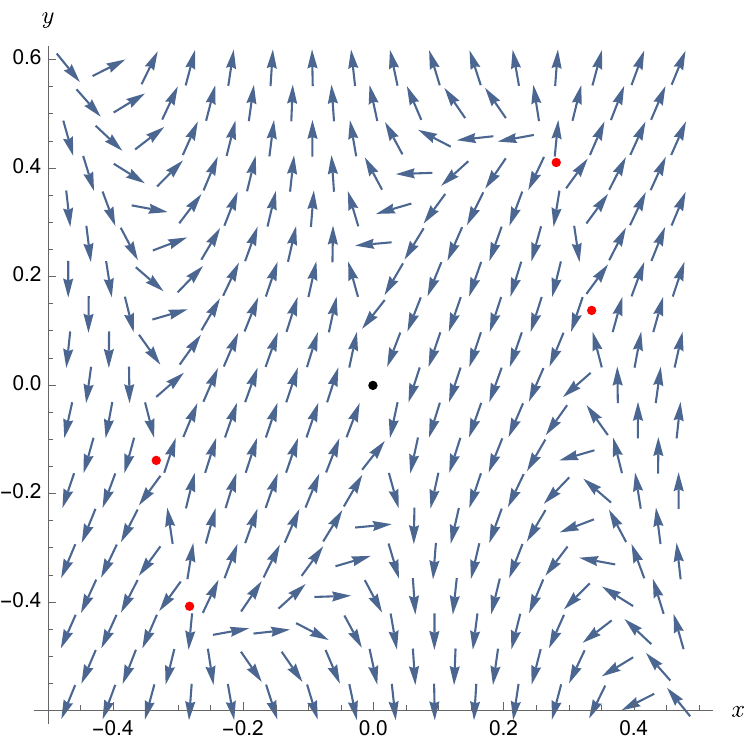}
         \caption{$N=6$}
         \label{fig:N=6}
     \end{subfigure}
     \hfill
     \begin{subfigure}[b]{0.3\textwidth}
         \centering
         \includegraphics[width=\textwidth]{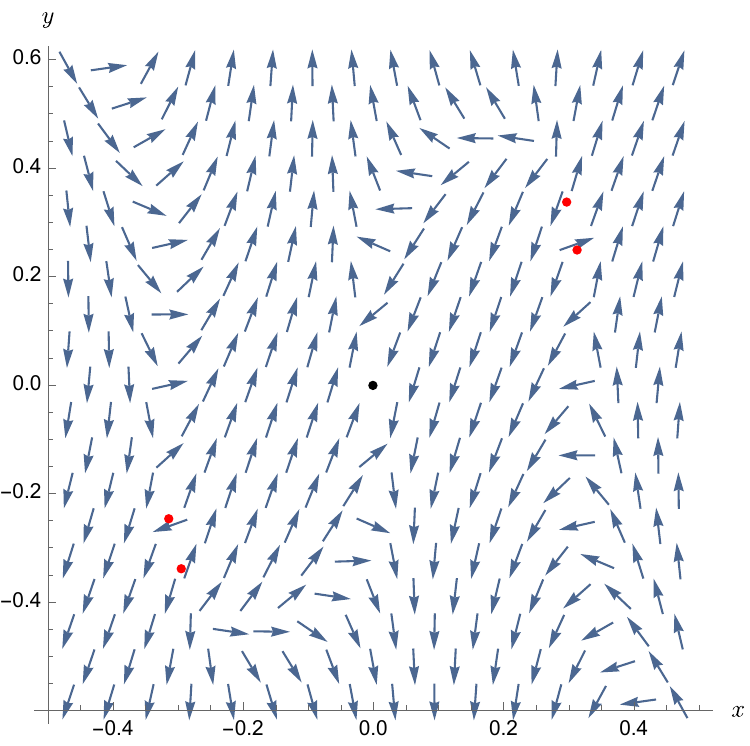}
         \caption{$N=7$}
         \label{fig:N=7}
     \end{subfigure}
        \caption{The real fixed points  and RG flows  in the $x\!-\!y$ plane for $2\le N\le 7$. The black dot denotes the trivial interface fixed points and the red dots denote the non-trivial ones.}
        \label{fig:six graphs}
\end{figure}

For $N=2$, all the fixed points are real. For $N=3$ and $4$, there is a pair of purely imaginary fixed points, and the rest are real. For $5\le N\le 7$, there are two pairs of  purely imaginary fixed points, and two pairs of non-trivial real fixed points. For $N\ge 8$, as we discussed above, the two pairs of real fixed points become complex, leaving only one real fixed point, i.e. the trivial one. Our results are reminiscent of those in \cite{Trepanier:2023tvb,Giombi:2023dqs,Raviv-Moshe:2023yvq} where an upper bound $N_{crit}=10$ was found for the existence of non-trivial real fixed points for the surface defects breaking
$O(N)$ to $O(N-1)$. However, the physics of our fixed points may be different: since we have to fine-tune several relevant operators (including the quadratic ones), we seem to have a multi-critical version of the normal (extraordinary) universality class.   
Altogether, the numbers of real/imaginary/complex non-trivial fixed points for each $N$ are summarized in table~\ref{countfixedpts}. We also plot RG flows between all the real fixed points for $2\le N\le 7$ in Figure~\ref{fig:six graphs}.

\begin{table}[h]
\centering
\begin{tabular}{ | c | c| c | c | c | c |} 
  \hline
  & $N=2$ & $N=3$ & $N=4$ & $5\le N\le 7$ & $N\ge 8$ \\ 
  \hline
Real & 6 & 6 & 4 & 4& 0 \\ 
  \hline
 Imaginary& 0 & 2 & 2 & 4&4\\ 
  \hline
   Complex& 0 & 0 & 0 & 0& 4  \\ 
  \hline
\end{tabular}
\caption{The number of \textbf{non-trivial} fixed points of each type for all integer $N\ge 2$ in the interface case.}
\label{countfixedpts}
\end{table}

For $N\geq 8$ there are no real fixed points, but there are IR stable fixed points with purely imaginary values of $g_1$ and $g_2$.\footnote{Such fixed points appear when $N$ becomes larger than 4.} For example, for $N=8$ we find a pair of purely imaginary fixed points
\begin{align}
(g_1^\star, g_2^\star)\approx \pm i\sqrt{\epsilon} \left(0.250457, -1.05133\right)
\end{align}
with positive critical exponents $\omega_+\approx 1.048\epsilon$ and $\omega_-\approx 0.07137\epsilon$.
While such theories are expected to be non-unitary, the path integral is well-defined. 

For large $N$, we can develop a $1/N$ expansion for these stable imaginary fixed points. More explicitly, we first solve the quartic equation $P_N(z)=0$. At the leading order in the large $N$ limit, $P_N(z)=0$ reduces to either $N z^3+N^2=0$ or $2 z^4+N z^3=0$. The first case includes a pair of unstable imaginary fixed points, and two pairs of complex fixed points, all of which have  exotic large $N$ behaviors. We will study them more carefully in appendix \ref{Exotic}.
 Here we  will only focus on the second case, which has a non-trivial solution $z = -\frac{N}{2}$. In other words, $P_N(z)$ has a solution $z^\star (N) = -\frac{N}{2}+\cdots$, where $\cdots$ denotes $1/N$ corrections. We compute these corrections up to $N^{-4}$:
 \begin{align}
z^\star(N) = -\frac{N}{2}-\frac{8}{N^2}+\frac{112}{N^3}+\CO\left(N^{-4}\right)~.
\end{align}
This solution leads to a pair of purely imaginary fixed points via \eqref{ztox2}
\begin{align}
x^\star(N)&=\pm i \left(\frac{1}{N} -\frac{3}{N^2} \right) +\mathcal{O}\left(N^{-3}\right) \, ,\crcr
y^\star(N)&= \mp i\left(\frac{1}{2}-\frac{3}{2N}+\frac{35}{4\, N^2}\right)  +\mathcal{O}\left(N^{-3}\right)  \, ,
\label{eq:stableFPlargeNint}
\end{align}
with positive critical exponents 
\begin{align}
\omega_+&=\epsilon\left(1+\frac{6}{N}-\frac{216}{N^2}\cdots \right)~,\nn\\
\omega_-&=\epsilon\left(\frac{1}{2}-\frac{11}{N}+\frac{252}{N^2}\cdots\right)~.
\end{align}
These two fixed points are thus stable in the IR.

\begin{figure}[t]
\centering
\includegraphics[width=7cm]{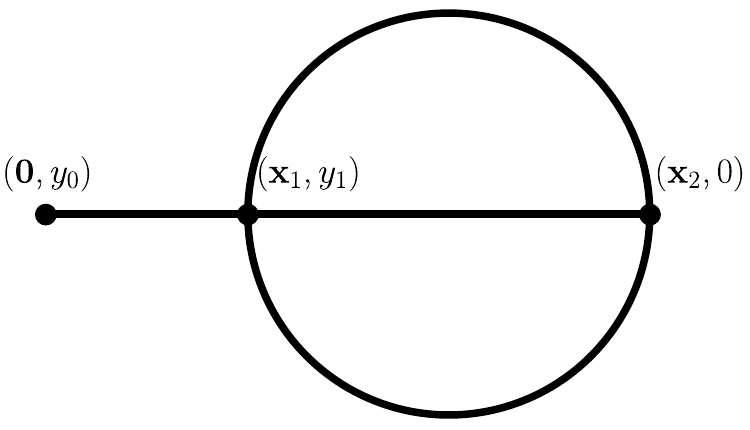}
\caption{The two-loop diagram contributing to the VEV of $\phi_N$.}
    \label{fig:1pt}
\end{figure}

The usual normal (extraordinary) transition in $d=3-\epsilon$ is characterized by the vacuum expectation value of the bulk operator behaving as 
$\langle \phi_N (y)\rangle  \sim \epsilon^{-1/2} |y|^{-1}$ \cite{Liendo:2012hy,Giombi:2020rmc}.
At the fixed points we have found, the VEV originates from the two-loop diagrams with the topology shown in Figure~\ref{fig:1pt}.  We compute these diagrams directly in position space 
\begin{align}
\langle\phi_N(y_0)\rangle = \frac{\lambda_4}{6}\left((N-1)\lambda_1+\lambda_2\right)\int_{\mathbb R^d}d^d\vec x_1\int_{\mathbb R^d}d^d\vec x_2\int_\mathbb{R} dy_1\frac{C_\phi^4}{(\vec x_1^2+y_{01}^2)^{\frac{d-1}{2}}(\vec x_{12}^2+y_1^2)^{3\frac{d-1}{2}}}~.
\end{align}
The integrals over $\vec x_1$ and $\vec x_2$ can be evaluated using 
\begin{align}
\int d^d\vec x\,\frac{1}{(\vec x^2+y^2)^\lambda}=\frac{ \pi^{\frac{d}{2}}\Gamma(\lambda-\frac{d}{2})}{\Gamma(\lambda)\left(y^2\right)^{\lambda-\frac{d}{2}}} ~.
\end{align}
The remaining $y_1$ integral is the one-dimensional version of \eqref{eq:bubble} with $\alpha = d-\frac{3}{2}$ and $\beta= -\frac{1}{2}$. Altogether, we get 
\begin{align}\label{phiN}
\langle\phi_N(y_0)\rangle = -\frac{\Gamma \left(d-\frac{5}{2}\right) \Gamma \left(\frac{d-1}{2}\right)^3\lambda_4\left((N-1)\lambda_1+\lambda_2\right)}{1536 \pi ^{d+\frac{3}{2}}  (d-2) \Gamma \left(\frac{3d-3}{2}\right)|y_0|^{2d-5}}~.
\end{align}
As $d\rightarrow 3$, since the $\Gamma$ function factors in \eqref{phiN} have a finite limit, we find:
\begin{align}
\langle\phi_N(y_0)\rangle \rightarrow - \frac{\lambda_4^\star\left((N-1)\lambda^\star_1+\lambda^\star_2\right)}{12 (4\pi)^4\,|y_0|}~.
\end{align}
Therefore, the VEV of $\phi_N$ is of order $\epsilon^{3/2}$ at the fixed points. Thus, we are finding a much milder effect in our multi-critical normal transition than in the usual normal transition.

\section{Scaling dimensions of quadratic critical operators}\label{quadO}

In this section, we compute the dimensions of the two $O(N-1)$ invariant quadratic operators $\mathbb{O}_1=\frac{\phi_a\phi_a}{\sqrt{N-1}}$ and $\mathbb{O}_2=\phi_N^2$.
To do so, we first compute the anomalous dimension matrix $\gamma$ for the mixing of the operators $\mathbb{O}_1$ and $\mathbb{O}_2$. The scaling dimensions will then be given by $2\Delta_{\phi}=d-1$ plus the eigenvalues of this matrix. 
To compute this matrix we follow the method of \cite{Fei:2014yja}. We denote $\mathbb{O}_i^{R}$ the renormalized operators. They can be expressed as:
\begin{align}
\mathbb{O}_1^{R}&=\mathbb{O}_1+\delta^{11}\mathbb{O}_1+\delta^{12}\mathbb{O}_2 \, , \crcr
\mathbb{O}_2^{R}&=\mathbb{O}_2+\delta^{21}\mathbb{O}_1+\delta^{22}\mathbb{O}_2 \, ,
\end{align}
where $\delta^{ij}$ are counterterms.
The mixing matrix is then defined by:
\begin{equation}
\gamma^{ij}=\mu \partial_{\mu} \left(-\delta^{ij}\right) \, .
\end{equation}

\begin{figure}[htbp]
\centering
\includegraphics[scale=1.5]{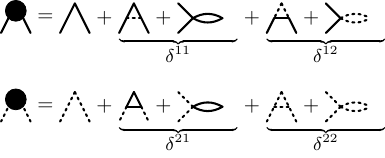}
\caption{Matrix elements of operators $\mathbb{O}_1$ and $\mathbb{O}_2$ at one-loop order. The solid lines represent a propagator between two fields $\phi_a\, , \,\, a=1,\dots N-1$ while the dashed lines represent a propagator between two fields $\phi_N$.}
\label{fig:mixing}
\end{figure}

We thus need to compute the counterterms $\delta^{ij}$. They are given by extracting the $\frac{1}{\epsilon}$ divergence from the graphs depicted in Figure~\ref{fig:mixing}, corresponding to the graphs $T$ and $B$ where one cubic vertex was replaced by a quadratic $\mathbb{O}_i$ vertex. Taking into account the factor $\sqrt{N-1}$ introduced in $\mathbb{O}_1$, we find for the matrix $\delta^{ij}$:

\begin{equation}
\delta^{ij}=\frac{\mu^{-\epsilon}}{\epsilon}\begin{pmatrix}
-\frac{g_1^2}{4}+\frac{N+1}{3}g_4 & -\frac{\sqrt{N-1}}{4}g_1^2+\frac{\sqrt{N-1}}{3}g_4 \\
-\frac{\sqrt{N-1}}{4}g_1^2+\frac{\sqrt{N-1}}{3}g_4 & -\frac{g_2^2}{4}+g_4 
\end{pmatrix} \, .
\end{equation}
We finally obtain for the mixing matrix:
\begin{equation}
\gamma^{ij}=\begin{pmatrix}
-\frac{g_1^2}{4}+\frac{N+1}{3}g_4 & -\frac{\sqrt{N-1}}{4}g_1^2+\frac{\sqrt{N-1}}{3}g_4 \\
-\frac{\sqrt{N-1}}{4}g_1^2+\frac{\sqrt{N-1}}{3}g_4 & -\frac{g_2^2}{4}+g_4 
\end{pmatrix} \, .
\end{equation}

When $N=1$, $\gamma^{ij}$ reduces to $g_4-\frac{g_2^2}{4}$.  So the scaling dimension of $\phi^2$ is $2-\frac{7}{6}\epsilon$ at the non-trivial fixed point \eqref{g2int} of the $N=1$ theory. For large $N$,  plugging in the stable imaginary fixed points \eqref{eq:stableFPlargeNint}, we obtain the following scaling dimensions  
\begin{equation}
\Delta_{\mathbb{O}_{-}}=2- \frac{\epsilon}{2} -\frac{2\,\epsilon}{N}\, , \quad \Delta_{\mathbb{O}_{+}}=2-\frac{5\epsilon}{N} \, ,
\end{equation}
corresponding to the operators:
\begin{align}
&\mathbb{O}_{-}=-N^{-\frac{1}{2}}\mathbb{O}_1+\mathbb{O}_2 \, ,\crcr
&\mathbb{O}_{+}=\left(N^{\frac{1}{2}}-\frac{1}{2}\,N^{-\frac{1}{2}}\right) \mathbb{O}_1+\mathbb{O}_2 \, .
\end{align}

\section{Interface in a model with symplectic fermions}
\label{sec:symp}

It is also interesting to consider continuations of the $O(N)$ models we have considered to negative $N$. The effective value of $N$ may be reduced by replacing two of
the commuting scalar fields, $\phi_1$ and $\phi_2$, by a pair of anticommuting scalars, $\theta$ and $\bar \theta$ (the latter are also known as symplectic fermions \cite{Kausch:2000fu,LeClair:2007iy}).
Let us study the model with $M$ pairs of symplectic fermions, $\theta_\alpha, \bar \theta_\alpha, \alpha=1, \ldots M$, along with one real scalar $\phi$:
\begin{align} \label{eq:action_symplectic}
		S[\phi, \theta]  \, &= \,  \int d^{d+1}x \, \bigg[ \frac{1}{2} \partial_{\mu}\phi\partial^{\mu}\phi + 
 \partial_{\mu}\theta_\alpha \partial^{\mu}\bar \theta_\alpha +
\frac{\lambda_4}{4!}\left(\phi^2+ 2\theta_\alpha \bar \theta_\alpha \right)^2 \bigg]   +  \int d^{d}\vec x \bigg[
		\lambda_1
		\phi \theta_a \bar \theta_\alpha +\frac{\lambda_2}{3!}\phi^3 \bigg] \, .
\end{align}
The quartic bulk action has $OSp(1|2M)$ symmetry, while the cubic interactions on the defect have the structure used in \cite{Fei:2015kta,Klebanov:2021sos}.
In general, the $OSp(1|2M)$ symmetry of the bulk theory
 is broken to $Sp(2M)$ by the defect or boundary. However, $M=1$ is the special case where the interaction on the defect preserves $OSp(1|2)$ if $\lambda_2=2 \lambda_1$, since $\frac{1}{3}\phi^3 +\phi\theta_1\bar\theta_1 = \frac{1}{3}(\phi^2+2\theta_1\bar\theta_1)^{\frac{3}{2}}$.
In this special case, as we will show explicitly below, there are indeed fixed points obeying this condition, as well as other fixed points where the defect breaks $OSp(1|2)$ to $Sp(2)$.  

The beta functions for the theory \eqref{eq:action_symplectic} can be obtained from those for the theory \eqref{eq:action_vectpr}
by replacing $N\rightarrow 1-2M$, and we find:
\begin{align}
\beta_{g_4}&=-\epsilon g_4 +\frac{g_4^2}{3}\left(9-2M\right) \, , \crcr
\beta_{g_1}&= -\frac{\epsilon}{2}g_1-\frac{1}{4}g_1^2\left(g_1+g_2\right) +\frac{g_4}{3}\left(2(3-M)g_1+g_2\right) \, , \crcr
\beta_{g_2}&=-\frac{\epsilon}{2}g_2-\frac{1}{4}\left(g_2^3-2M\,g_1^3\right)+g_4\left(3g_2-2Mg_1\right) \, .
\end{align}
Since at the one-loop fixed point $g^\star_4= 3\epsilon/(9-2M)$, to keep the quartic coupling positive, we need to take $M\leq 4$.
Once $g_4$ is positive, real cubic couplings do not seem to cause problems for the convergence of the path integral.

In the case $M=1$, the fixed point for the quartic coupling becomes:
\begin{equation}
g_4^{\star}=\frac{3\epsilon}{7}+\mathcal{O}(\epsilon^2) \, .
\end{equation}
For the cubic couplings, we find, besides the trivial fixed point, four pairs of real fixed points 
\begin{align}
g_1^{\star} &=  \pm \sqrt{\frac{10\,\epsilon}{21}} , \quad g_2^{\star}=  \pm 2\sqrt{\frac{10\,\epsilon}{21}} \, , \crcr
g_1^{\star} &\approx   \pm 0.65067\sqrt{\epsilon}, \quad g_2^{\star} \approx  \pm 0.604968\sqrt{\epsilon} \, , \crcr
g_1^{\star} &\approx   \pm 0.897903\sqrt{\epsilon}, \quad g_2^{\star} \approx  \mp 1.9905\sqrt{\epsilon}\, ,  \crcr
g_1^{\star} &\approx   \pm 1.3832\sqrt{\epsilon} , \quad g_2^{\star} \approx \mp 1.67773\sqrt{\epsilon} \,.
\label{eq:fpsymp}
\end{align}
The first one preserves $OSp(1|2)$ with $g_2^{\star}=2 g_1^{\star}$. 
The only IR stable fixed point is the trivial one; all other fixed points have at least one relevant direction. The corresponding critical exponents are given by:
\begin{align}
\omega_+& \approx -0.690476 \epsilon , \quad \omega_- \approx -0.714286\epsilon  \, , \crcr
\omega_+& \approx 0.50253 \epsilon, \quad \omega_-\approx -0.434223\epsilon  \, , \crcr
\omega_+& \approx 0.352249 \epsilon, \quad \omega_- \approx -2.17771\epsilon  \, , \crcr
\omega_+& \approx \left(-0.76428 -0.600305 i\right) \epsilon, \quad  \omega_- =\omega_+^* \, .
\end{align}
The fixed point with $g_2^{\star}=2 g_1^{\star}$ is relevant in both directions. Two pairs of fixed points have one relevant and one irrelevant direction. The last fixed point has complex critical exponents with negative real part and $\omega_+=\omega_-^*$.

For $2\leq M \leq 4$, all non-trivial fixed points are complex with both real and imaginary parts non-zero. Again the only stable fixed point is the trivial one. For the complex fixed points, the critical exponents are complex and at least one has a negative real part.

For generic $M$, we can analyze the fixed points using the same approach as in section \ref{sec:vector model}. We make the substitution $ g_1 =\sqrt{8\,\epsilon}\, x, \,  g_2 =\sqrt{8\,\epsilon}\, y$ in the beta functions. This leads to the following quartic equation in $z=\frac{y}{x}$:
\begin{equation}
P_{M}(z)= 2z^4+(3-2M)z^3-(9+2M)z^2+(9-6M)z+2M(2M+3)=0 \, ,
\end{equation}
and $x$ is related to $z$ by
\begin{equation}
x^2=\frac{1}{4(9-2M)}\left(\frac{1-2M}{1+z}+2 \right) \, .
\end{equation}

The discriminant of $P_{M}(z)$ is positive for $1\leq M \leq 4$. For $M=1$ it has four real roots, leading to the real fixed points of \eqref{eq:fpsymp}. For $2 \leq M \leq 4$, $P_{M}(z)$ has two pairs of complex conjugate roots, leading to complex fixed points. For $M\geq 4$, the discriminant of $P_{M}(z)$ is negative: $P_{M}(z)$ has two real roots and two complex conjugate roots. The real roots then lead to purely imaginary fixed points.

\section{Cubic interactions at the boundary}
\label{sec:boundary}

We consider the $O(N)$ vector model in the half space $y\ge 0$, with the same cubic interactions as in section \ref{sec:vector model} inserted on the boundary $y=0$. The action is given by 
\begin{align} \label{eq:action_boundary}
		S[\phi]  \, &= \,  \int_{y\ge 0} d^{d+1}x \, \bigg[ \frac{1}{2} \partial_{\mu}\phi_I\partial^{\mu}\phi_I + \frac{\lambda_4}{4!}\left(\phi_I \phi_I \right)^2 \bigg]   +  \int d^{d}\vec x \bigg[
		\frac{\lambda_1}{2} \, 
		\phi_N \phi_a \phi_a +\frac{\lambda_2}{3!}\phi_N^3 \bigg] \, .
\end{align}
Imposing the Neumann boundary conditions, the free bulk propagator becomes
\begin{align}
\langle\phi_I(\vec x_1, y_1)\phi_J(\vec x_2, y_2)\rangle_B=\delta_{IJ}\left(\frac{C_\phi}{(\vec x^2_{12}+(y_1-y_2)^2)^{\frac{d-1}{2}}}+\frac{C_\phi}{(\vec x^2_{12}+(y_1+y_2)^2)^{\frac{d-1}{2}}}\right)~,
\end{align}
where $C_\phi$ is defined in \eqref{interprop}. Performing Fourier transformations for $\vec x_1$ and $\vec x_2$, we obtain 
\begin{align}
\langle\phi_I(\vec p_1, y_1)\phi_J(\vec p_2, y_2)\rangle_B=(2\pi)^d\delta^d(\vec p_1\!+\!\vec p_2)\,\delta_{IJ} \frac{e^{-|\vec p_1||y_{1}-y_2|}+e^{-|\vec p_1|(y_{1}+y_2)}}{2|\vec p_1|}~,
\end{align}
from which we can easily read off the boundary-to-bulk and purely boundary propagators
\begin{align}\label{bprop}
K_{IJ}^{(B)} (\vec p, y)= \frac{e^{-|\vec p| y}}{|\vec p|}\delta_{IJ} , \,\,\,\,\,\,\, G_{IJ}^{(B)}(\vec p) = \frac{1}{|\vec p|}~. 
\end{align}
Compared to the interface case,  the propagators in \eqref{bprop} are larger by a factor of 2, and $y$ is valued in $\mathbb R_+$ instead of $\mathbb R$. For this reason, the diagram $T$ in Figure~\ref{fig:1_2_loops} should be multiplied by a factor of $8$ because it has three boundary propagators, and the diagram $B$ should be twice bigger because it contains two boundary-to-bulk propagators and one $y$ integral. Another new feature compared to the interface case is that the field $\phi_I$ picks up a one-loop {\it boundary} anomalous dimension from the snail diagram with the bulk vertex \cite{gompper1985conformal, McAvity:1995zd, Giombi:2020rmc}:\footnote{We are grateful to Simone Giombi for pointing out the importance of this one-loop effect in the boundary case.} 
\begin{align}
\gamma_{\hat \phi} = -\frac{(N+2)\lambda_4}{6(4\pi)^2 }=-\frac{(N+2) g_4}{6 }.
\end{align}
It is not hard to reproduce this result using our methods. The one-loop correction in momentum space is 
\begin{align}
G^{(1)}_2(\vec p) = - \frac{(N+2)\lambda_4}{6}\, \int\frac{d^d\vec k}{(2\pi)^d}\int_0^\infty d y\,  \left(\frac{e^{- |\vec p|y}}{|\vec p|}\right)^2\frac{1+e^{-2 |\vec k| y}}{2|\vec k|}~.
\end{align}
The $y$ integral yields the sum of $\frac{1}{2|\vec k||\vec p|}$ and $\frac{1}{2|\vec k|(|\vec p|+|\vec k|)}$. Only the latter can contribute to the wavefunction renormalization.
Its $d$ dimensional integral over $\vec k$ can be easily evaluated  in spherical coordinates, and we find 
\begin{align}
G^{(1)}_2(\vec p) = \frac{(N+2)\lambda_4}{3(4\pi)^2|\vec p|}\frac{1}{\epsilon}+\mathcal{O}(1) \, .
\end{align}
Defining the wavefunction renormalization $Z_\phi$ through $\phi_{I}^{(0)} = Z_\phi^\frac{1}{2} \phi_I$ with $Z_\phi = 1+\delta_\phi$, the corresponding counterterm is then $\delta G_2 (\vec p) = -\frac{\delta_\phi}{|\vec p|}$.  
In the minimal subtraction scheme, we choose $\delta_\phi = \frac{(N+2)\lambda_4}{3(4\pi)^2 \epsilon}$, leading to the boundary anomalous dimension of $\phi$
\begin{align}
\gamma_{\hat \phi} = -\frac{1}{2} \lambda_4\partial_{\lambda_4}\left(\epsilon \delta_\phi\right) =-\frac{(N+2)\lambda_4}{6(4\pi)^2 }~.
\end{align}

Due to the appearance of the one-loop boundary anomalous dimension, the boundary beta functions $\beta^{(B)}_{g_i}$ pick up the contributions $3 g_i \gamma_{\hat \phi}= - \frac{(N+2) }{2 } g_i  g_4$. Thus, instead of the beta functions
\eqref{betag} found for the interface, we now find
\begin{align}
\beta^{(B)}_{g_1}& = -\frac{\epsilon}{2}g_1 -2\,g_1^2\left(g_1+g_2\right) + \frac{2}{3}g_4\big ((N+5)g_1+g_2\big ) -\frac{(N+2) }{2 }  g_1 g_4  \, ,  \crcr
\beta^{(B)}_{g_2}&= -\frac{\epsilon}{2}g_2 -2\left((N-1)g_1^3+g_2^3\right) + 2 g_4 \big ( (N-1)g_1+3g_2 \big )   -\frac{(N+2) }{2 } g_2 g_4 \, ,
\end{align}
where we have used the rescaling defined in \eqref{gdef}.

Altogether, the one-loop $\beta$ functions in this boundary theory become
\begin{align}\label{betagboundary}
\beta_{g_4}&= -\epsilon g_4 +\frac{N+8}{3}g_4^2  \, , \crcr
\beta^{(B)}_{g_1}&=-\frac{\epsilon}{2}g_1 -2\,g_1^2\left(g_1+g_2\right) + \frac{N+14}{6}\, g_4g_1+ \frac{2}{3}g_4 g_2 \, ,  \crcr
\beta^{(B)}_{g_2}&= -\frac{\epsilon}{2}g_2 -2\left((N-1)g_1^3+g_2^3\right) +2(N-1)g_4g_1-\frac{N-10}{2} g_4 g_2    \, .
\end{align}
If we make a further rescaling $\tilde g_{1,2}= \sqrt{8} g_{1,2}$, we then get
\begin{align}\label{newbetagboundary}
\beta^{(B)}_{\tilde g_1}&=-\frac{\epsilon}{2}\tilde g_1 -\frac{1}{4} \,\tilde g_1^2\left(\tilde g_1+\tilde g_2\right) + \frac{N+14}{6}\, g_4 \tilde g_1+ \frac{2}{3}g_4 \tilde g_2\, ,  \crcr
\beta^{(B)}_{\tilde  g_2}&= -\frac{\epsilon}{2}\tilde g_2 -\frac{1}{4} \left((N-1)\tilde g_1^3+\tilde g_2^3\right) +2(N-1)g_4\tilde g_1-\frac{N-10}{2} g_4 \tilde g_2  \, .
\end{align}
Now the only difference from the beta functions \eqref{betag} in the interface theory is that the terms linear in the bulk coupling $g_4$ have different coefficients.
The theory with a free bulk is then the same as for the interface. 
In the case of an interacting bulk, the critical value of the  quartic coupling is again given by \eqref{eq:quarticWF}. Substituting this value into $\beta^{(B)}_{\tilde g_1}$ and $\beta^{(B)}_{\tilde g_2}$, we find that they become
\begin{align}
&\beta^{(B)}_{\tilde g_1}= \frac{3\epsilon}{N+8}\tilde g_1+\frac{2\epsilon}{N+8}\tilde g_2 -\frac{1}{4} \,\tilde g_1^2\left(\tilde g_1+\tilde g_2\right)~,\nn\\ 
&\beta^{(B)}_{\tilde  g_2}= \frac{(11-2N)\epsilon}{N+8}\tilde g_2+\frac{6(N-1)\epsilon}{N+8}\tilde g_1-\frac{1}{4} \left((N-1)\tilde g_1^3+\tilde g_2^3\right)  ~.
\end{align}

\begin{figure}[h]
     \centering
     \begin{subfigure}[b]{0.3\textwidth}
         \centering
         \includegraphics[width=\textwidth]{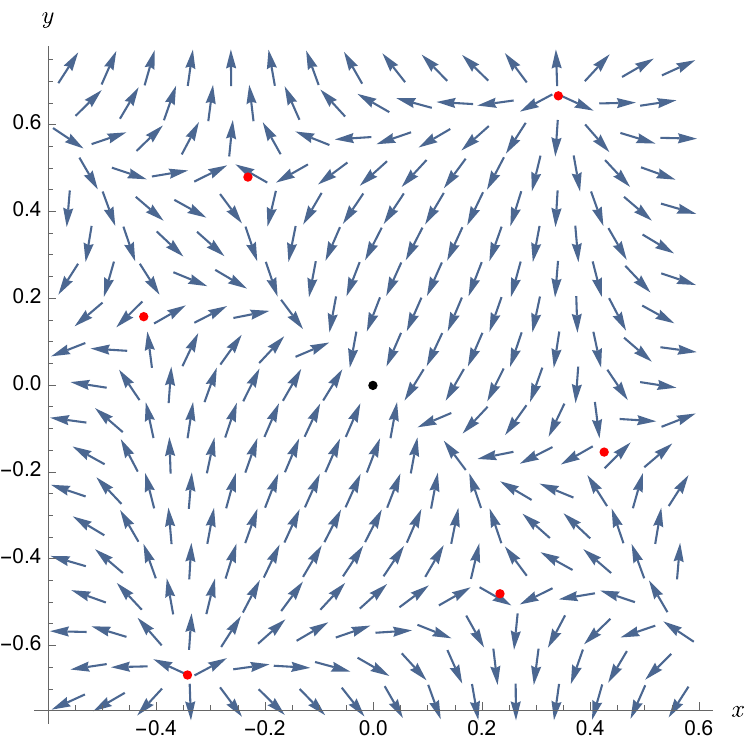}
         \caption{$N=2$}
         \label{fig:BN=2}
     \end{subfigure}
     \hfill
     \begin{subfigure}[b]{0.3\textwidth}
         \centering
         \includegraphics[width=\textwidth]{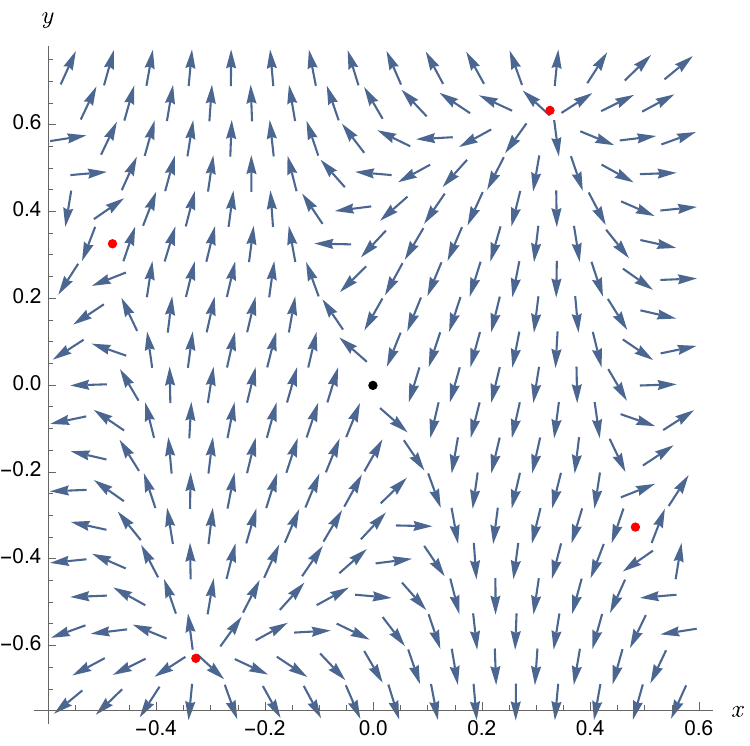}
         \caption{$N=3$}
         \label{fig:BN=3}
     \end{subfigure}
     \hfill
     \begin{subfigure}[b]{0.3\textwidth}
         \centering
         \includegraphics[width=\textwidth]{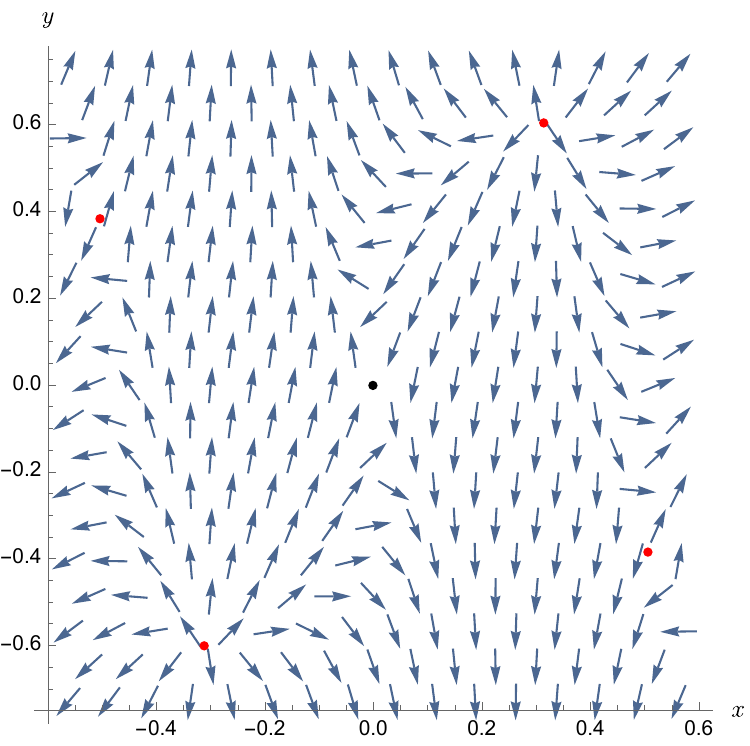}
         \caption{$N=4$}
         \label{fig:BN=4}
     \end{subfigure}
          \begin{subfigure}[b]{0.3\textwidth}
         \centering
         \includegraphics[width=\textwidth]{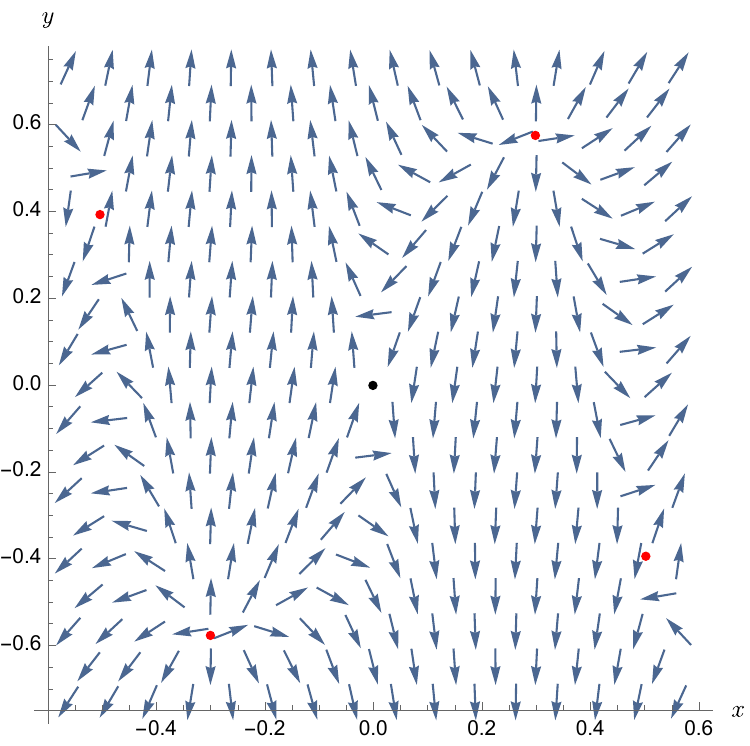}
         \caption{$N=5$}
         \label{fig:BN=5}
     \end{subfigure}
     \hfill
     \begin{subfigure}[b]{0.3\textwidth}
         \centering
         \includegraphics[width=\textwidth]{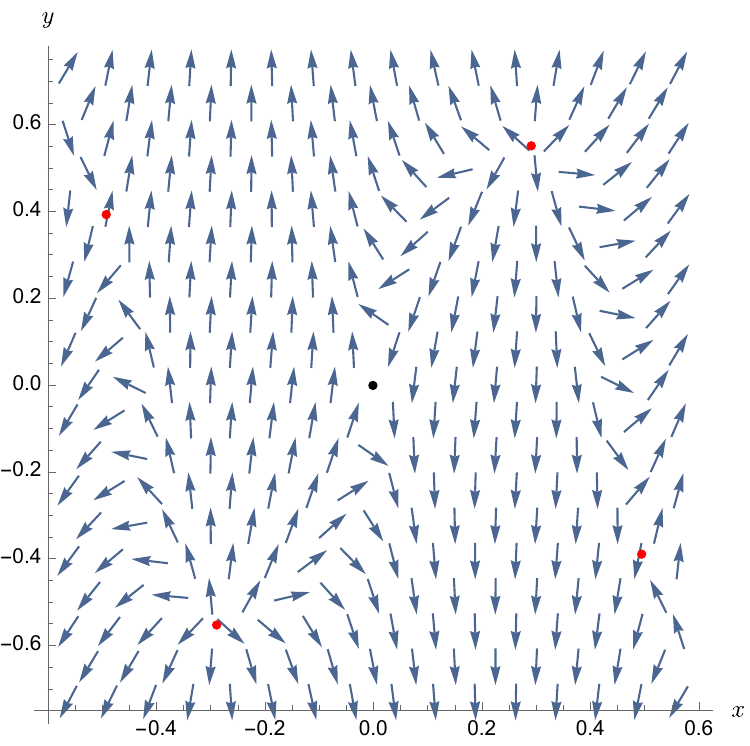}
         \caption{$N=6$}
         \label{fig:BN=6}
     \end{subfigure}
     \hfill
     \begin{subfigure}[b]{0.3\textwidth}
         \centering
         \includegraphics[width=\textwidth]{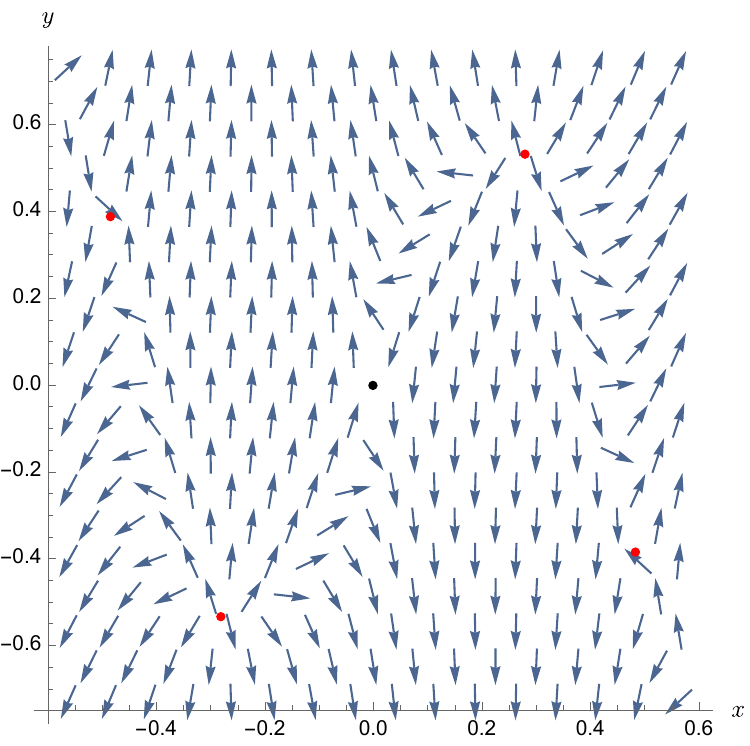}
         \caption{$N=7$}
         \label{fig:BN=7}
     \end{subfigure}
        \caption{The real fixed points  and RG flows  in the boundary case for $2\le N\le 7$.  The black dot denotes the trivial fixed point and the red dots denote the non-trivial ones.}
        \label{fig:six Bgraphs}
\end{figure}

For $N=1$, we find two real non-trivial fixed points for the cubic couplings:
\begin{equation}
\gt_2^{\star}=\pm 2\sqrt{\,\epsilon}+\mathcal{O}(\epsilon^{3/2}) \, ,
\end{equation}
with critical exponent $\omega_2 = - 2\epsilon$. So the cubic operator $\phi^3$ has scaling dimension $\Delta_{\phi^3} = 3-3\epsilon$.
For $N\ge 2$, we can analyze the fixed points using the same approach as in section \ref{sec:vector model}. To briefly recap the method, we make the substitution $\tilde g_1 =\sqrt{8\,\epsilon}\, x, \tilde g_2 =\sqrt{8\,\epsilon}\, y$ in the beta functions, which leads to a quartic equation of $z = \frac{y}{x}$:
\begin{align}\label{B4}
P_N^{(B)}(z)\equiv 2 z^4+3 z^3+(2N-11) z^2-(2N+7) z-3 (N-1) =0~,
\end{align} 
and a simple relation between $x$ and $z$ 
\begin{align}
x^2 = \frac{1}{2(N+8) }\left (\frac{1}{z+1}+2\right)~.
\end{align}
The quartic function of $z$ has a critical value $N_{\rm crit}\approx 2.50495$, which is a zero of the corresponding discriminant. It has another critical value $N'_{\rm crit} = \frac{5}{2}$, which corresponds to the sign change  of $\frac{1}{z+1}+2$ at  certain root of $P_N^{(B)}(z)$. For $N=2$, there are 6 non-trivial  fixed points, and they are all real. 
For $2<N<N_{\rm crit}'$, in addition to the six real fixed points, there is also a pair of imaginary fixed points. When $N$ crosses $N_{\rm crit}'$ from below but is still smaller than 
$N_{\rm crit}$, a pair of real fixed points become imaginary. In other words, in the tiny region $N_{\rm crit}'<N< N_{\rm crit}$, there are 4 non-trivial real fixed points and 4 imaginary fixed points. When $N$ crosses $N_{\rm crit}$ from below, the imaginary fixed points become complex. The $4$ real fixed points remain and they are all unstable: two of them have one relevant direction, and the other two have two relevant directions. We summarize the properties of all the non-trivial fixed points for integer $N\ge 2$ in table~\ref{countfixedpts2}. Interestingly, there are real fixed points for all values of $N\geq 2$. We also plot the RG flows between the real fixed points for $2\le N\le 7$ in Figure~\ref{fig:six Bgraphs}.

\begin{table}[h]
\centering
\begin{tabular}{ | c | c|c|} 
  \hline
  & $N=2$ & $N > 2$ \\ 
  \hline
Real & 6 & 4  \\ 
  \hline
 Imaginary& 0 & 0 \\ 
  \hline
   Complex& 0 & 4  \\ 
  \hline
\end{tabular}
\caption{The number of \textbf{non-trivial} fixed points of each type for all integer $N\ge 2$ in the boundary case.}
\label{countfixedpts2}
\end{table}

The quartic equation \eqref{B4} has two types of large $N$ solutions: two real solutions and two complex solutions. This leads to two pairs of real fixed points and two pairs of complex fixed points. The leading large $N$ behaviors of the real fixed points are determined by the quadratic equation $2z^2-2 z-3=0$. We compute the $1/N$ corrections up to $N^{-2}$:
\begin{align}
&x^{\star}_1=\pm \sqrt{\frac{5+\sqrt{7}}{2}}\frac{1}{N^{1/2}}\mp\sqrt{\frac{39059+8287\sqrt{7}}{14}}\frac{1}{8N^{3/2}} +\mathcal{O}(N^{-2})\, , \crcr
&y^{\star}_1= \mp \frac{\sqrt{13-\sqrt{7}}}{2N^{1/2}}\pm \sqrt{\frac{9379+137\sqrt{7}}{7}}\frac{3}{16N^{3/2}}  +\mathcal{O}(N^{-2}) \, ,
\end{align}
and 
\begin{align}
&x^{\star}_2=\pm \sqrt{\frac{5-\sqrt{7}}{2}}\frac{1}{N^{1/2}}\mp\sqrt{\frac{39059-8287\sqrt{7}}{14}}\frac{1}{8N^{3/2}}  +\mathcal{O}(N^{-2}) \, , \crcr
&y^{\star}_2= \mp \frac{\sqrt{13+\sqrt{7}}}{2N^{1/2}}\pm \sqrt{\frac{9379-137\sqrt{7}}{7}}\frac{3}{16N^{3/2}}  +\mathcal{O}(N^{-2}) \, .
\end{align}
The first pair has one relevant and one irrelevant direction, and the second pair has two relevant directions. The corresponding critical exponents are:
\begin{align}
\omega_{+}&=\epsilon\left(\frac{4\sqrt{2}}{N}\left(7+5\sqrt{7}\right)+\mathcal{O}(N^{-2}) \right)\, , \crcr
\omega_{-}&=\epsilon\left(-4\sqrt{2}-\frac{3\sqrt{2}}{N}\left(11+5\sqrt{7}\right)+\mathcal{O}(N^{-2})\right) \, ,
\end{align}
and 
\begin{align}
\omega_{+}&=\epsilon\left(- \frac{4\sqrt{2}}{N}\left(5\sqrt{7}-7\right)+\mathcal{O}(N^{-2}) \right) \, , \crcr
\omega_{-}&=\epsilon\left( -4\sqrt{2}+\frac{3\sqrt{2}}{N}\left(5\sqrt{7}-11\right)+\mathcal{O}(N^{-2}) \right)\, .
\end{align}

The calculation of the VEV of $\phi_N$ is analogous to that in the interface case, and the two-loop diagram \ref{fig:1pt} again makes the leading contribution. The result in \eqref{phiN} should be multiplied by a factor of 8 in the boundary case, because each boundary-to-bulk propagator contains an extra factor of 2 and the mirror image in the bulk-to-bulk propagator can be removed by extending $y_1$ to the whole real line. In $D=4-\epsilon$ the VEV is again found to be of order $\epsilon^{3/2}$.

So far we have discussed the $O(N)$ model with various positive values of $N$. In order to adapt the discussion to boundaries in the $OSp(1|2M)$ models, we can replace $N$ by $1-2M$.
For example, for $M=1$, we find a special $OSp(1|2)$ symmetric fixed point with 
\begin{align}
g_2^{\star}=2 g_1^{\star}= \pm \sqrt{\frac{2\epsilon}{3}}\, .
\end{align}
This is an unstable fixed point with two relevant directions.
There are also fixed points where the symmetry is broken to $Sp(2)$
\begin{align}
&g_1^\star \approx \pm 0.332179 \sqrt{\epsilon},\quad g_2^\star \approx \mp 1.06191\sqrt{\epsilon} \, ,\nn\\
&g_1^\star  \approx \pm 0.434673 \sqrt{\epsilon},\quad g_2^\star \approx \pm  0.23906 \sqrt{\epsilon} \, ,\nn\\
&g_1^\star  \approx \pm 0.793277\sqrt{\epsilon},\quad g_2^\star \approx \mp 0.67679\sqrt{\epsilon}~.
\end{align}
For other values of $M$, there are only the fixed points that break the $OSp(1|2M)$ symmetry of the bulk theory down to $Sp(2M)$.

\section*{Acknowledgments}

We are grateful to Simone Giombi, Alexander S\"{o}derberg-Rousu, and Shimon Yankielowicz for useful discussions. We also thank Simone Giombi and Max Metlitski for very useful comments on a draft of this paper.
This work was supported in part by the US National Science Foundation under Grant No.~PHY-2209997. SH is grateful for the support of the Princeton Center for Theoretical Science where part of this work was carried out. Nordita is supported in part by NordForsk.

\appendix
\section{Exotic Large $N$ limits}\label{Exotic}
\subsection{Interface with a free bulk}
Let us first discuss the fixed points with $\lambda_4=0$. We denote the  solutions of $z^3-z^2-z +N-1=0$ (c.f. \eqref{xzeq}) by $z_\alpha(N)$ with $z_0(N)$ being real and $z_{1,2}(N)$ being complex. The corresponding $x$ and $y$ are then given by:
\begin{align}\label{xN}
x_{\alpha, \pm} (N) =\pm  \frac{z_\alpha(N)-1}{2\sqrt{N-2}}, \,\,\,\,\, y_{\alpha,\pm} (N) =  \pm\frac{z_\alpha(N) \left(z_\alpha(N)-1\right)}{2\sqrt{N-2}}~.
\end{align}
At large $N$, these solutions  are:
\begin{align}\label{allZ}
&z_0(N)=-N^{\frac{1}{3}}+\frac{1}{3}-\frac{4}{9\,N^{\frac{1}{3}}}\,+\frac{38}{81\,N^{\frac{2}{3}}}\,-\frac{152}{729\,N^{\frac{4}{3}}}  +\mathcal{O}(N^{-\frac{5}{3}})\, , \nn\\
&z_{1} (N)=z_2(N)^* = e^{\frac{i\pi}{3}} N^{\frac{1}{3}}+\frac{1}{3}+\frac{4\, e^{-\frac{i\pi}{3}}}{9\, N^{\frac{1}{3}}}-\frac{38\,e^{\frac{i\pi}{3}}}{81\,N^{\frac{2}{3}}}  +\frac{152\,e^{-\frac{i\pi}{3}} }{729\,N^{\frac{4}{3}}}  +\mathcal{O}(N^{-\frac{5}{3}})~.
\end{align}
Since $z_2(N)$ is simply the complex conjugate of $z_1(N)$, we will only focus on $z_0(N)$ and $z_1(N)$. Plugging them into \eqref{xN}, we obtain the corresponding fixed points
\begin{align}
&x_{0,\pm}=\mp\left(\frac{1}{2\,N^{\frac{1}{6}}}  + \frac{1}{3\, N^{\frac{1}{2}}}+\frac{2}{9\, N^{\frac{5}{6}}} + \frac{43}{162\, N^{\frac{7}{6}}}  \right)+\mathcal{O}(N^{-\frac{3}{2}}) \, , \nn\\
 &y_{0,\pm}=\pm\left(\frac{N^{\frac{1}{6}}}{2} +\frac{1}{6\,N^{\frac{1}{6}}} +\frac{1}{3 \,N^{\frac{1}{2}}} +\frac{17}{162\, N^{\frac{5}{6}}}+ \frac{91}{486\, N^{\frac{7}{6}}}   \right)+\mathcal{O}(N^{-\frac{3}{2}})  \, ,
\end{align}
and 
\begin{align}
x_{1,\pm}&=\pm\left(\frac{e^{\frac{i \pi }{3}}}{2\, N^{\frac{1}{6}}}-\frac{1}{3\,N^{\frac{1}{2}}}+\frac{2 \, e^{-\frac{i\pi}{3}}}{9\, N^{\frac{5}{6}}}+  \frac{43\, e^{\frac{i\pi}{3}}}{162\, N^{\frac{7}{6}}} \right)+\mathcal{O}(N^{-\frac{3}{2}}), \crcr
y_{1,\pm}&= \mp\left(\frac{e^{-\frac{i \pi }{3}}}{2} N^{\frac{1}{6}}+\frac{e^{\frac{i \pi }{3}}}{6\,N^{1/6}}-\frac{1}{3N^{\frac{1}{2}}} +\frac{17\,e^{-\frac{i \pi }{3}}}{162\, N^{\frac{5}{6}}} + \frac{91 e^{\frac{i \pi }{3}}}{486\, N^{\frac{7}{6}}}\right)+\mathcal{O}(N^{-\frac{3}{2}})\, .
\end{align}
Plugging $z_0(N)$ and $z_1(N)$ into \eqref{omegapmfree} yields the critical exponents for the real fixed points
\begin{align}
\omega_-=\epsilon\left(-\frac{3}{2} N^{\frac{1}{3}}-2-\frac{8}{3\,N^{\frac{1}{3}}}-\frac{53}{27\,N^{\frac{2}{3}}}-\frac{8}{3N}+\mathcal{O}(N^{-\frac{4}{3}})\right) , \quad\omega_+=\epsilon  \,,
\end{align}
and for the complex fixed points 
\begin{align}
\omega_-&=\epsilon\left(\frac{3\, e^{\frac{i\pi}{3}}}{2} N^{\frac{1}{3}}-2+\frac{8\, e^{-\frac{i\pi}{3}}}{3\, N^{\frac{1}{3}}}+\frac{53\,e^{\frac{i\pi}{3}}}{27\,N^{\frac{2}{3}}}-\frac{8}{3N}+\mathcal{O}(N^{-\frac{4}{3}})\right) ,\quad
\omega_+=\epsilon~ .
\end{align}
The critical exponent $\omega_-$ blows up in both cases.

\subsection{Interface with an interacting bulk}

Next, we tune $\lambda_4$ to its critical value. In this case, we need to solve  $P_N(z)=0$ (c.f. \eqref{z4}) in the large $N$ limit. As discussed at the end of section \ref{Intint}, the  quartic equation $P_N(z)=0$ has 
 two types of large $N$ solutions. In this appendix, we are interested in solutions whose leading large $N$ behavior is determined by the cubic equation  $Nz^3+N^2=0$. These solutions are 
\begin{align}
&z_{0} (N)= -N^{\frac{1}{3}}-\frac{1}{3}-\frac{16}{9 \, N^{\frac{1}{3}}}+\CO\left(\frac{1}{N^{\frac{2}{3}}}\right)\nn~ ,\\
&z_{1} (N)=z_2(N)^*= e^{\frac{i\pi}{3}}N^{\frac{1}{3}}-\frac{1}{3}-\frac{16\, e^{-\frac{i\pi}{3}}}{9 \, N^{\frac{1}{3}}}+\CO\left(\frac{1}{N^{\frac{2}{3}}}\right)~,
\end{align}
where $z_0(N)$ leads to a pair of purely imaginary fixed points, and $z_{1,2} (N)$ correspond to two pairs of complex fixed points.

To compare with the purely imaginary fixed points given by \eqref{eq:stableFPlargeNint}, we solve the fixed points corresponding to $z_0(N)$
\begin{align}
&x_{0, \pm} (N) = \pm\frac{i}{2}\left(\frac{1}{N^{\frac{1}{6}}}+\frac{1}{3 N^{\frac{1}{2}}}\right)+\mathcal{O}\left(N^{-\frac{5}{6}}\right)~,\nn\\
&y_{0, \pm} (N) = \mp \frac{i}{2}\left(N^{\frac{1}{6}}+\frac{2}{3 N^{\frac{1}{6}}}+\frac{1}{6\, N^{\frac{1}{2}}}\right)+\mathcal{O}\left(N^{-\frac{5}{6}}\right) \, .
\end{align}
They have  critical exponents 
\begin{align}
&\omega_+ = \epsilon\left(\frac{3}{2} N^{\frac{1}{3}}+2+\frac{1}{N^{\frac{1}{3}}}+\cdots\right) ~,\nn\\
&\omega_- = \epsilon \left(-1+\frac{2}{3 N^{\frac{1}{3}}}+\cdots\right)~,
\end{align}
and hence are unstable. $\omega_+$ also shows some exotic large $N$ behavior.

\,

\,

\bibliographystyle{JHEP-3}

\bibliography{Refs} 

\addcontentsline{toc}{section}{References}


\end{document}